\documentclass[%
 reprint,
 amsmath,amssymb,
 aps,
]{revtex4-2}

\usepackage{graphicx}% Include figure files
\usepackage{dcolumn}% Align table columns on decimal point
\usepackage{bm}% bold math
\usepackage{hyperref}
\begin{document}

\preprint{APS/123-QED}

\title{Simulating the DFT Algorithm for Audio Processing}% Force line breaks with \\
\thanks{More precisely, the usefulness of linear algebra for the same.}%

\author{Omkar Deshpande}
  \email{omkar.d@ahduni.edu.in}%Lines break automatically or can be forced with \\
\author{Kharanshu Solanki}%
 \email{kharanshu.s@ahduni.edu.in}
 
\author{Sree Pujitha Suribhatla}%
 \email{sree.s@ahduni.edu.in}
\affiliation{%
 Department of Physics, School of Arts and Sciences, Ahmedabad University, Ahmedabad - 380009\\
}%

\author{Sanya Zaveri}
  \email{sanya.z@ahduni.edu.in}%Lines break automatically or can be forced with \\
\author{Luv Ghodasara}%
 \email{luv.g1@ahduni.edu.in}
\affiliation{%
 Department of Comptuter Science, School of Arts and Sciences, Ahmedabad University, Ahmedabad - 380009\\
}%
\date{\today}% It is always \today, today,

\begin{abstract}
Since the evolution of digital computers, the storage of data has always been in terms of discrete bits that can store values of either 1 or 0. Hence, all computer programs (such as MATLAB), convert any input continuous signal into a discrete dataset. Applying this to oscillating signals, such as audio, opens a domain for processing as well as editing. The Fourier transform, which is an integral over infinite limits, for the use of signal processing is discrete. The essential feature of the Fourier transform is to decompose any signal into a combination of multiple sinusoidal waves that are easy to deal with. The discrete Fourier transform (DFT) can be represented as a matrix, with each data point acting as an orthogonal point, allowing one to perform complicated transformations on individual frequencies. Due to this formulation, all the concepts of linear algebra and linear transforms prove to be extremely useful here. In this paper, we first explain the theoretical basis of audio processing using linear algebra, and then focus on a simulation coded in MATLAB, to process and edit various audio samples. The code is open ended and easily expandable by just defining newer matrices which can transform over the original audio signal. Finally, this paper attempts to highlight and briefly explain the results that emerge from the simulation. 
\end{abstract}

\maketitle

%\tableofcontents

\section{\label{sec:level1}Introduction\protect}

The Fourier transform, in the most basic sense, is used as a tool to break down a waveform into its composite amplitudes and frequencies. It has the ability to construct any repetitive function of time or a waveform as a combination of individual sinusoidal vibrations, which are individually functions of time.\\

Although Joseph Fourier first used the notion of a Fourier transform to explain certain scenarios pertaining to heat transfer [1], it now finds its uses in a huge number of fields. These include practical applications like signal processing, optics, image processing, as well as theoretical applications like calculating eigenvalues of unitary operators in quantum algorithms.\\

Fourier transform finds its most crucial application in the field of signal processing, i.e., to study a given signal which is a waveform (a function of time), and then obtain useful and important information from these waveforms. Waveforms can be the range of any function changing with time - from electromagnetic waves to sound waves and waves produced in strings to electrical signals.\\ 

Since signals and waveforms are so essential and are present everywhere in the world around us, it is necessary that one understands signal processing and the role of the Fourier transform as an important tool to analyze it. More precisely computers are equipped to implement the discrete Fourier transform, which deals with discrete numbers of equally spaced samples, rather than continuous systems.\\ 

This paper aims to analyze an audio form by converting it into a waveform and then applying the DFT algorithm to it, via a coded simulation. Once the waveform is analyzed, the code obtains the number of highest amplitudes or frequencies that constitute the signal. Further, we try to demonstrate how and why these composite amplitudes are important in various fields.  These objectives are to be carried out with the aid of the MATLAB programming languge, which provides simple and helpful functions to carry out the same.

\section{\label{sec:level2}Theory and Significance\protect}
\subsection{\label{sec:level2}Discrete Data and Digital Signals}

Data can be broadly classified into discrete and continuous sets of data. If a piece of data is defined to be between two particular real values, and further, if the data can be anything between these two points, then it can be called as continuous data. On the contrary, if the data can only be defined such that there is a finite gap on either side of it and it cannot be taken as anything else, it would be defined as discrete.\\

An illustration of this is the binary of analog signal and digital signal. Digital signals are the discrete quantities which convey information regarding something with respect to a predefined set of values. For example, the binary codes used in computers can only read “on” (true or 1) and “off” (false or 0). On the other hand, an analog signal is a continuous signal with one of the quantities changing with time (like temperature or humidity). Theoretically, analog signals are meant to have infinitely many values and the data can take any value within the infinitesimal range of another value. For example, a light bulb can have not only “on" or “off” but it can be described as “dimmer” or “brighter” on a relative scale [2].\\

Signals or waveforms are usually continuous data, and although Fourier transform is an important tool to analyze them, our computers are not equipped to perform these functions which are continuous in nature. Our computers work in the “on/off” system, which means that they can only deal with discrete data points. So then, how do we deal with continuous data? Well, digital signal processing takes care of it. In digital signal processing, a physical or a natural signal which is continuous, gets sampled and quantized [3]. The signal’s values are picked out at equal intervals of time and only those data points are considered essential to study the signal. The digital signal is an ordered code obtained from a finite set of values.\\
\begin{figure}[hbt!]
\includegraphics[width=50mm]{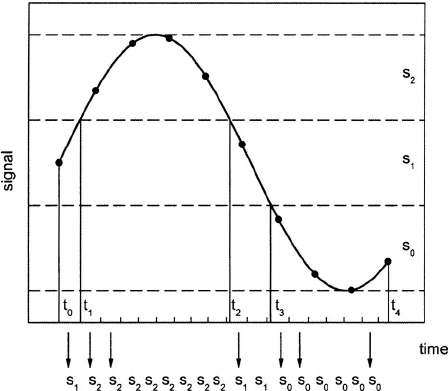}% Here is how to import EPS art
\caption{\label{fig:epsart} A digital signal picks out values which are spaced out in equal intervals of time.}
\end{figure}

If one understands this, then one can understand how the Fourier transform and discrete Fourier transform (DFT) would work. The Fourier transform is essentially an integral transform; it deals with continuous data from a waveform. However, a DFT will only pick out a specific required number of sample points from the available data, and then analyze them and give out crucial information, regarding the signal that we originally had [4]. Therefore, the DFT is a vital tool in signal processing and we shall explore the use of it in later sections, once we understand the working of Fourier transform itself.

\subsection{\label{sec:citeref}Integral Transforms}
The Fourier transform is basically a special case of a more general kind of representation of functions. This general class of representations is known as an integral transform. What is an integral transform? Well, the moment we see the term transform, it implies some kind of a change. When we say that we are applying a Fourier transform (or in general any integral transform) on a given function, we mean that we are changing from one particular kind of function to another.\\

Let’s say we have a function $f(t)$, which is dependent on the variable $t$, and another function $g(x)$, which is dependent on the variable $x$.  An integral transform then gives us a way to interchange (or transform) between these two functions, by means of an integral over a specific interval. In terms of linear algebra, we could say the function $f(t)$ is a vector (or a set of vectors) in the vector space $V$ with basis $t$, and the function $g(x)$ is a vector (or a set of vectors) in the vector space $W$ with basis $x$. In this case, and integral transform would be analogous to a linear transformation $T:V\rightarrow W$ (or rather a change of basis from t to x), that would tell us how to transform from $f(t)$ to $g(x)$, and vice-versa. The vice-versa case is usually called an inverse transformation, or an inverse integral transform. The functions $f(t)$ and $g(x)$ are related through an integral transform as,

 \begin{equation}
g(x)=\int_{a}^{b} f(t) K(x, t) d t
\end{equation}

So, we get the function $g(x)$ by multiplying the function $f(t)$ by another function $K(x,t)$ (which is dependent on both $x$ and $t$), and then integrating this product w.r.t. $t$, over the interval $[a,b]$.Here, $a$ and $b$ are constants, and $K$ is also called the kernel function. We can also say that $g(x)$ itself is the integral transform of the given function $f(t)$. For the sake of notation, it is useful to introduce a linear operator,

\begin{equation}
g(x)=\int_{a}^{b} K(x, t) d t=T[f(t)]
\end{equation}

The values of $a$ and $b$, and the form of $K(x,t)$ is different for different functions. This gives us a compact form of the integral transform as,

\begin{equation}
T[f(t)]=\int_{a}^{b} f(t) K(x, t) d t \mid
\end{equation}

It turns out that the integral transform operator $(T)$ is in fact a linear operator, as it has the following properties of linearity:
\begin{itemize}
  \item  \textbf{Closure under addition}: The integral transform of a sum of two functions($f_1$ and $f_2$) is equal to the sum of the integral transforms of the two functions individually.
    \begin{equation}
T\left[f_{1}(t)+f_{2}(t)\right]=T\left[f_{1}(t)\right]+T\left[f_{2}(t)\right] 
\end{equation}
  \item \textbf{Closure under scalar multiplication}: The integral transform of a scalar multiple of a function $f(t)$ is equal to the scalar $c$ times the integral transform of the function.
  \begin{equation}
\begin{array}{c}
T[c f(t)]=c T[f(t)] 
\end{array}
\end{equation}
\end{itemize}

Now, the kind of integral transforms that are useful and have varied applications (like the Fourier transform) are generally invertible. In terms of the linear operator notation, this means that $T^-1$ exists, and usually there’s a convenient method to evaluate it. So, given that,

\begin{equation}
g(x)=T[f(t)]
\end{equation}

there exists a $T^-1$ such that,

\begin{equation}
T^{-1}[g(x)]=f(t)
\end{equation}

Apart from the Fourier transform, other kinds of integral transforms include the Laplace transform, the Hankel transform and the Mellin transform.\\

Why is it that the Fourier transform (or any kind of integral transform) is used a lot for solving both theoretical, and real-world practical problems? Well, any kind of mathematical tool is generally introduced when solving a problem becomes an exceedingly difficult task, or when solving the problem becomes much easier while using the mathematical tool in a certain context. For example, it is often easier to calculate spherical volumes (via volume integrals) using the spherical coordinate system rather than the Cartesian one. It is easier to use the Lagrangian or Hamiltonian formulation to analyze certain complex dynamical systems rather than the Newtonian one [5]. It is easier to decouple Maxwell’s inhomogeneous equations using the Lorenz gauge than using the Coulomb gauge [6]. Symmetry principles are so often used in physics to simplify massive calculations. The same is the case with the Fourier transform. It is quite a sophisticated mathematical tool, which makes seemingly difficult problems, easy to solve.\\

\begin{figure}[hbt!]
\includegraphics[width=50mm]{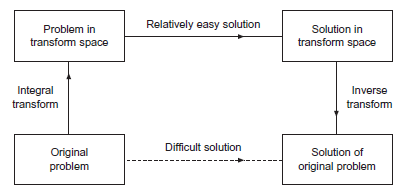}% Here is how to import EPS art
\caption{\label{fig:epsart} The integral transform algorithm [7]}
\end{figure}

The integral transform algorithm is very well explained by Arfken, Weber and Harris [7] using the following diagram (Figure 2), which is self-explanatory. 

\subsection{\label{sec:level2}The Fourier Transform}
The Fourier transform can also be viewed as an extension of the idea of a Fourier series. The Fourier transform is to a Fourier series, what a tensor is to a vector. In other words, the Fourier transform is a sort of generalization of the concept of the Fourier series.\\

The Fourier series is basically a way to represent functions that have some periodicity, i.e., it is a way to represent periodic functions. The important thing to note about the Fourier series is that the periodic functions are defined over a finite period. For a given function $f(t)$, its Fourier series expansion is given as the sum of a constant and a series of sinusoids (sines and cosines). 

\begin{equation}
f(t)=\frac{a_{0}}{2}+\sum_{n=1}^{\infty}\left[a_{n} \cos \left(\frac{n \pi t}{L}\right)+b_{n} \sin \left(\frac{n \pi t}{L}\right)\right]
\end{equation}

So, what if we want to represent functions with an infinite period (which basically implies an aperiodic function)? In that case, we use the Fourier transform. We saw how an integral transform takes a function in one kind of space (or domain) to another. The Fourier transform specifically takes a function from the time domain to another function in the frequency domain. How is this important or useful? Well, as we will see, this has an interesting application in signal processing and frequency analysis of signals. For a given function $f(t)$, its Fourier transform is given as,

\begin{equation}
F(\omega)=\frac{1}{\sqrt{2 \pi}} \int_{-\infty}^{\infty} f(t) e^{i \omega t} d t
\end{equation}

The only requirement or constraint on the function $f(t)$ is that the integral of the absolute value of $f(t)$ must be a finite quantity, i.e., $\int_{-\infty}^{\infty}|f(t)| d t$ is finite. 

\subsection{\label{sec:citeref}Application of the Fourier Transform to Audio Decomposition}

So far, we have discussed the basic notion of a Fourier transform and some of its properties. However, we haven’t yet touched upon how linear algebra is used along with the Fourier transform to perform audio decomposition algorithms or frequency analysis of signals.\\

Let’s start with thinking about what an audio waveform or a signal is. Generally, an audio file will contain a single wave with varying amplitudes and intensities with time. Therefore, we can think of any audio file (such as a .mp3 file or a .wav file) as intensity fluctuations in time, or simply a function of time. 

\begin{figure}[hbt!]
\includegraphics[width=50mm]{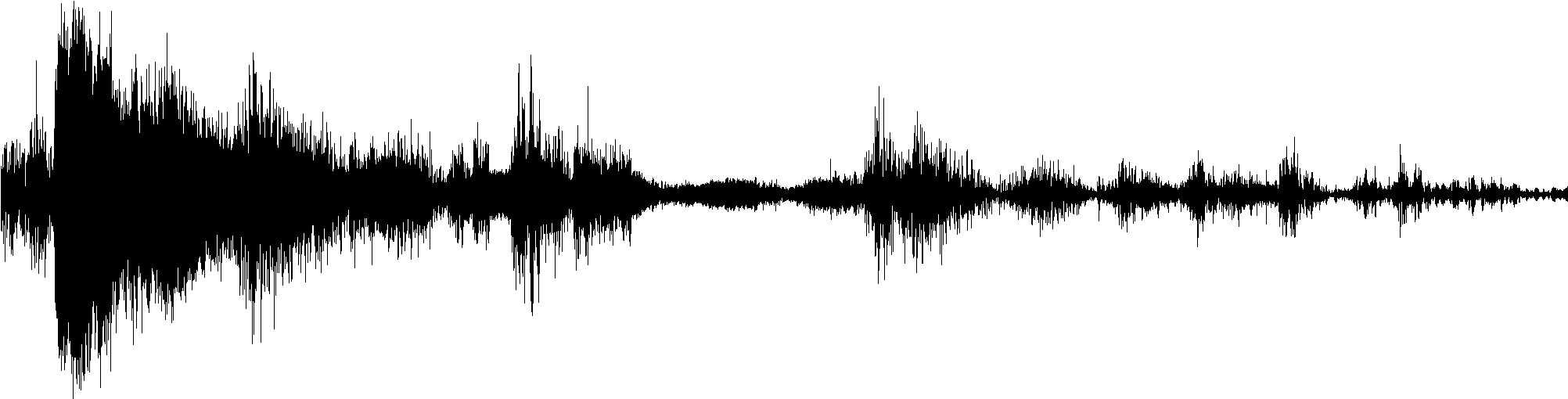}% Here is how to import EPS art
\caption{\label{fig:epsart} An audio waveform as a combination of different frequencies.}
\end{figure}

Now, what use is a Fourier transform here? Well, a majority of the times, it is the case that a single audio waveform is a combination of different waves of different frequencies, rather than being a single sinusoidal wave of a specific frequency. What if we want to find out what different kinds of frequencies have been combined to give the final audio? The Fourier transform comes in at this point. By applying Fourier transform on the audio waveform, we can basically know which individual frequencies the audio is made up of. But what is the use of this? Why would someone want to know what kind of frequencies an audio waveform is made up of? Audio artists, or video editors (and other people as well) find it useful to do so. Why? Well, consider that a music composer is given an audio waveform, which has a disturbing high pitch frequency to it, which is annoying to listen to. If the composer wants to remove this high frequency disturbance in the audio, then he/she would first want to know what different kinds of frequencies are individually contributing to the final audio waveform. To do so, he/she can apply a Fourier transform to the audio waveform, which then tells him/her what is that high frequency which is causing an annoying disturbance; and then he/she can remove that high frequency contribution from the audio waveform!\\

We still don’t know where linear algebra comes into play. Well consider this. An average human (or even a superhuman probably) would not be able to listen to an audio and create its waveform by hand! Computing a Fourier transform for that waveform would be another mammoth task to do by mind. Therefore, computational methods are required to convert an audio into its waveform, and then to apply a Fourier transform on the waveform. The thing with computers though, is that they are not continuous systems (like the Fourier transform), but they work in discrete units called bits (0s and 1s). Therefore, it would be impossible to perform the ordinary continuous Fourier transform algorithm using a computer. This is where we introduce an incredibly important algorithm known as the Discrete Fourier Transform (DFT) algorithm. And linear algebra comes in handy in the DFT algorithm [8]. The fundamental process of the DFT algorithm is that of approximating a Fourier series approximation of the audio waveform on a finite period where the waveform is periodic. So, what we are actually approximating is the Fourier series expansion, and not the infinite Fourier transform. The DFT actually leads to one of the most important algorithms - the Fast Fourier Transform (FFT) algorithm - which has applications in image compression, audio compression, and other highly advanced computer applications. For now, let us focus on the DFT. The reason why linear algebra is an important ingredient of the DFT algorithm, is because the DFT is a mathematical transformation that can be written in terms of a huge matrix multiplication (the matrix is called the DFT matrix). \\

So, we talked about approximating functions that are periodic using infinite sums of sines and cosines, using the Fourier series. But in most cases, we don’t necessarily have an analytic (or continuous) function to deal with. In computer related algorithms, we usually have to deal with some sort of measurement data or with discrete data sets from a simulation or an experiment. So, what we have is a function $f$ defined at discrete locations $x_0,x_1,x_2...x_n$, as shown below in the figure.

\begin{figure}[hbt!]
\includegraphics[width=50mm]{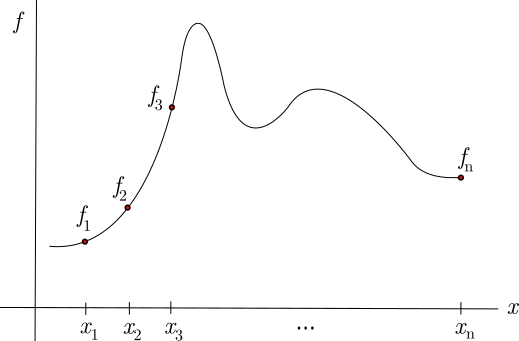}% Here is how to import EPS art
\caption{\label{fig:epsart} A function $f$ defined at discrete locations $x_0,x_1,x_2...x_n$}
\end{figure}

These discrete data or function values $f_0, f_1, f_2, ... , f_n$ at the respective locations $x_1,x_2,x_3...x_n$, can be represented as a vector of data as follows.

\begin{equation}
f=\left[\begin{array}{c}
f_{0} \\
f_{1} \\
f_{2} \\
\vdots \\
f_{n}
\end{array}\right]
\end{equation}

Now, what we want to do is compute the DFT of this vector of data, i.e., we want to break the data up into a sum of sines and cosines. And this comes in handy while analyzing audio. If the data is an audio data, then we can use the DFT to figure out what sort of tones add up to make that data. Further, based on the frequency content of the data, we can diagnose whether something is wrong with the audio (like some unwanted deviations or spikes in the data). So, in a nutshell, turning a data vector into its sinusoidal components using the DFT algorithm, can be very useful. The important point to press is that a DFT is just a Fourier series expansion of discrete data, rather than a continuous, analytic function. Recall how while expanding a continuous function using the Fourier series, we took the function and turned them into the Fourier coefficients $a_n$ and $b_n$, that were multiplied to the sines and cosines. With the DFT, we are essentially doing the same thing. We take our data $\hat{f}$, and obtain its Fourier transform vector, , which is given as,

\begin{equation}
\hat{f}=\left[\begin{array}{c}
\hat{f}_{0} \\
\hat{f}_{1} \\
\hat{f}_{2} \\
\vdots \\
\hat{f}_{n}
\end{array}\right]
\end{equation}

The Fourier transform vector of the data, consists of frequency components of the data (because recall that a Fourier transform takes a function from the time domain to the frequency domain). So, in the $\hat{f}$ vector, the first frequency component $\hat{f}_{0}$ tells us how much of that first low-frequency is in the data $f$, the second component  $\hat{f_1}$ tells us how much of the second (slightly higher) frequency is there in the data, and so on till the $n$th frequency (which is the highest possible frequency). If we let $k=0,1,2,3,...,n$, then we can calculate the $k$th Fourier coefficient, or the $k$th frequency component as,

\begin{equation}
\hat{f}_{k}=\sum_{j=0}^{n-1} f_{j} e^{-\frac{i 2 \pi j k}{n}}
\end{equation}

We will get to what the exponential term means in a while, but this is essentially how we calculate the Fourier coefficients or frequency components. For now, it is important to keep in mind that if we have data, we can get its Fourier transform, but if we have a Fourier transform, we should also be able to reconstruct the original data. This sort-of inverse DFT is given as,

\begin{equation}
f_{k}=\frac{1}{n} \sum_{j=0}^{n-1} \widehat{f}_{j} e^{\frac{i 2 \pi j k}{n}}
\end{equation}

All this basically means that if we have the data, and we apply the DFT to it, and get the Fourier coefficients. Each of these Fourier coefficients literally just tell us how much of each specific frequency is required to reconstruct the original data from the coefficients. An interesting thing to note is that all of the terms in a DFT are multiplied in the form of an exponential, which is an integral multiple of $e^{-2\pi i/n}$. This term actually defines a fundamental frequency, given as,

\begin{equation}
\omega_{n}=e^{-2 \pi i / n}
\end{equation}

This fundamental frequency is related to what kind of sines and cosines can be approximated with $n$ discrete values in the domain $x$ (refer figure 3). In other words, $\omega_n$ is the kind of fundamental frequency that we need to work with, if we have an interval of n data points. Now, all of these Fourier transforms are adding up integer multiples of that fundamental frequency times the data values (and the same with the inverse Fourier transform). So, we can use this fundamental frequency to compute a matrix, which when multiplied by the data, would give the corresponding Fourier transform (or Fourier coefficients).\\

Computing the DFT for each data point can be a tedious task by hand, leave alone creating an entire computer program that does that! Instead, what we do with the DFT, is that we try to write what the sum would be for each $k$, in terms of a matrix operation. This matrix, called the DFT matrix, has components which are in terms of the fundamental frequency $\omega_n$. It can be shown that this is exactly analogous to the infinite series of sines and cosines that we get with the ordinary Fourier series. Now, for the zeroth frequency $(f_0)$, then all of the exponents in the DFT would be $e^0$, i.e., 1. So, all the data points will have a unit coefficient. This gives us a linear equation of the form,

\begin{equation}
f_{0}+f_{1}+f_{2}+\ldots+f_{n}=\hat{f}_{0}
\end{equation}

Similarly, for $k=1$, we get the equation, 

\begin{equation}
1 f_{0}+\omega_{n} f_{1}+\omega_{n}^{2} f_{2}+\ldots+\omega_{n}^{n-1} f_{n}=\hat{f}_{1}
\end{equation}

If we find the equation for each particular $k$-value, then we find that we can write these linear equations in the matrix representation as,

\begin{equation}
\left[\begin{array}{c}
\hat{f}_{0} \\
\hat{f}_{1} \\
\hat{f}_{2} \\
\vdots \\
\hat{f}_{n}
\end{array}\right]=\left[\begin{array}{ccccc}
1 & 1 & 1 & \cdots & 1 \\
1 & \omega_{n} & \omega_{n}^{2} & \cdots & \omega_{n}^{n-1} \\
1 & \omega_{n}^{2} & \omega_{n}^{4} & \cdots & \omega_{n}^{2(n-1)} \\
\vdots & \vdots & \vdots & \ddots & \vdots \\
1 & \omega_{n}^{n-1} & \omega_{n}^{2(n-1)} & \cdots & \omega_{n}^{(n-1)^{2}}
\end{array}\right]\left[\begin{array}{c}
f_{0} \\
f_{1} \\
f_{2} \\
\vdots \\
f_{n}
\end{array}\right]
\end{equation}

Here, note that the fundamental frequencies are all complex numbers (which can be easily shown using Euler’s formula). So, the DFT matrix is a complex matrix, and all the Fourier coefficients on the LHS of the matrix equation are complex valued. 

\subsection{\label{sec:citeref} Applications of the DFT matrix}

Now that we have understood how Fourier Transforms and DFT work, we can delve deeper into the applications of DFT and the DFT matrix.  The Fourier Transform vector,

\begin{equation}
\hat{f}=\left[\begin{array}{c}
\hat{f}_{0} \\
\hat{f}_{1} \\
\hat{f}_{2} \\
\vdots \\
\hat{f}_{n}
\end{array}\right]
\end{equation}

which we obtained in the previous section is known as the frequency spectrum; it tells us about the contribution of the individual frequencies of the original waveform.\\

Vital information is stored in the sinusoids which form a signal. Our universe is an encompassment of many oscillating objects, and understanding these signals is important. For example, all the sounds that we hear are a result of various vibrations and combinations of different sound waves; we are able to speak because of the vibration produced in our vocal cords; there is a periodic movement in the working of a water turbine, etc. All of these things give out signals which are waveforms and DFT is essential to understand the frequency, amplitude, and phase of the constituent sinusoidal waves [9].\\

This DFT algorithm is used vastly in our project to filter out the ingredient frequencies of our considered audio data. This can further be used to de-noise signals or modify them in any way. However, note that the FFT (fast Fourier transform) would be better to use in case the data set is extremely huge, because that’s what an FFT basically is. The FFT is just an efficient method to calculate the DFT for large datasets.

\section{Audio Processing Samples}

\subsection{The Audio Processing Code Implemented in MATLAB}

The MATLAB code file can be found here:\\
\url{https://drive.google.com/file/d/10ndp1fgEiaBCc_0osNp_-ihh_AL6ZyRt/view?usp=sharing}. \\

The MATLAB code is designed with a user interface which is easily expandable as one needs it. It will first take in an audio file name as an input, import the audio file and create its waveform along with its Fourier transform. Then, one can perform a list of tasks with it. The working of each task is explained as comments in the program.\\

The code has incorporated certain ways to calculate multiplication and transforms of matrices such that the RAM usage as well as the performance optimized for regular use. After performing various operations on the matrix, the user can then choose to do the inverse transform which will give back the transformed matrix in time-dependent domain. The task ‘plots’, creates the plot of the edited waveform as well as its Fourier counterpart that can be observed as graphs. Additional functions to save the edited sample as well as to listen to it (multiple times) are added for secondary observations.\\

\subsection{Results from the Simulation}

\subsubsection{DFT of a Simple Chord Progression}

The program takes in a sound waveform which is a superposition of different frequencies and looks as shown below in figure 5.

\begin{figure}[hbt!]
\includegraphics[width=50mm]{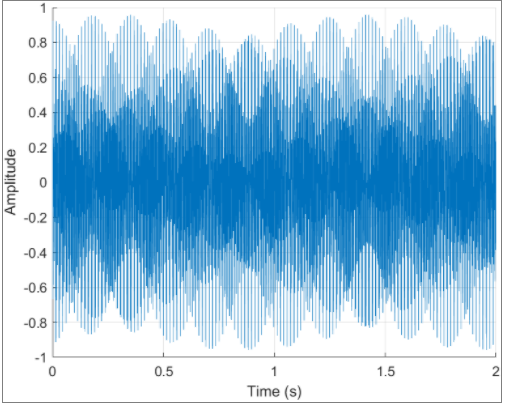}% Here is how to import EPS art
\caption{\label{fig:epsart} The amplitudes of all the frequencies which constitute the audio.}
\end{figure}

When the DFT function is applied onto it, we obtain the following graph:

\begin{figure}[hbt!]
\includegraphics[width=50mm]{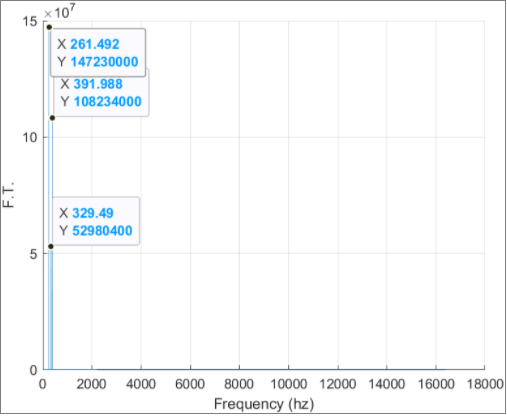}% Here is how to import EPS art
\caption{\label{fig:epsart}  Frequencies of individual waves which constitute the audio wave}
\end{figure}

Once the DFT is applied on the original sound, it processes and decomposes it; thus, yielding individual frequencies of the different waves which combine to form the original wave. In Figure 6, we see that we obtain three different frequencies which can be identified with their respective musical notes.\\

The standard note to frequency table relevant to our obtained frequencies is as shown below.

\begin{table}[h]%The best place to locate the table environment is directly after its first reference in text
\caption{\label{tab:table1} Standard notes and their frequencies [10].}
\begin{ruledtabular}
\begin{tabular}{lcdr}
\textrm{Note}&
\textrm{Frequency}
\\
\colrule
$C_4$ & 261.63\\
$E_4$ & 329.63\\
$G_4$ & 392.00\\
\end{tabular}
\end{ruledtabular}
\end{table}

Therefore, we can obtain a rough correlation between our obtained frequencies and the standard frequencies. We have 261.49 which pertain to the $C_4$ note, then 329.49 pertaining to $E_4$ and 391.98 pertain to $G_4$. As the basic laws of music theory says, it corresponds to a $C$-major chord which could be heard from the original signal.

\subsubsection{DFT of Complex Music Signals}

The sound wave in the previous section of composed of two different audios whose waveforms are shown below:

\begin{itemize}
  \item  \textbf{Sample: Every Tear Drop is a Waterfall by Coldplay.}
  
\begin{figure}[hbt!]
\includegraphics[width=50mm]{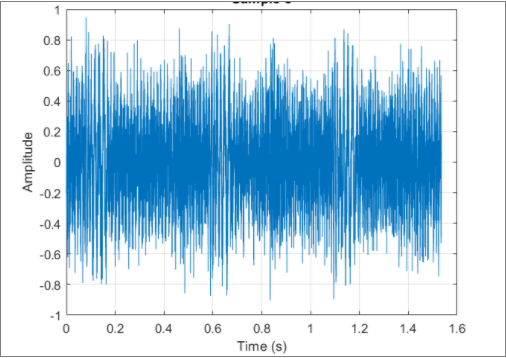}% Here is how to import EPS art
\caption{\label{fig:epsart}   The amplitudes of all the frequencies which constitute the audio.}
\end{figure}

\begin{figure}[hbt!]
\includegraphics[width=50mm]{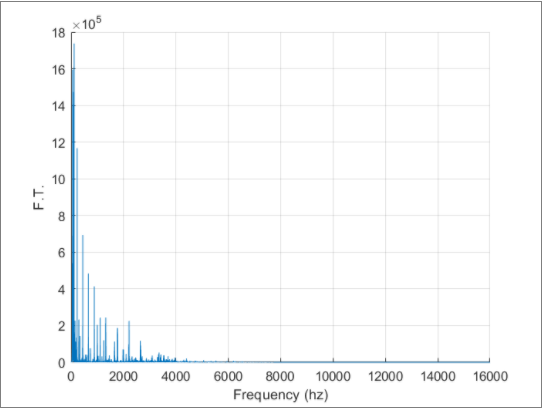}% Here is how to import EPS art
\caption{\label{fig:epsart}   Frequencies of individual waves which constitute the audio wave}
\end{figure}
  
  \item \textbf{Sample- Master of Puppets by Metallica:}
  
\begin{figure}[hbt!]
\includegraphics[width=50mm]{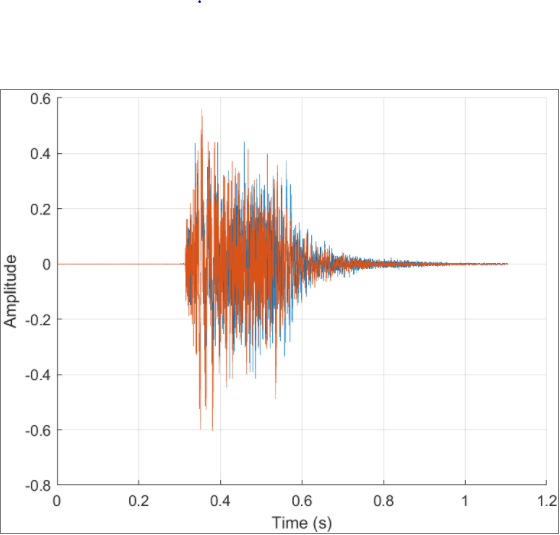}% Here is how to import EPS art
\caption{\label{fig:epsart}   The amplitudes of all the frequencies which constitute the audio.}
\end{figure}

\begin{figure}[hbt!]
\includegraphics[width=50mm]{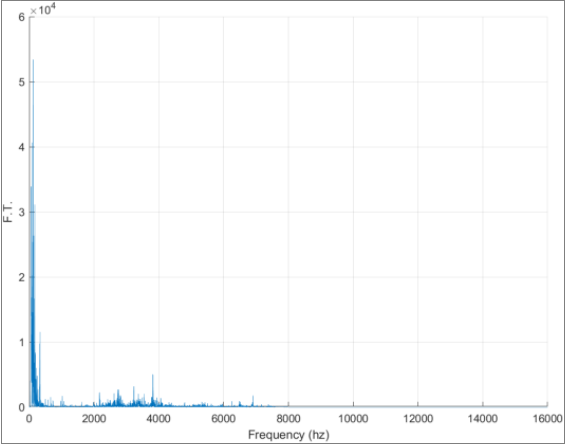}% Here is how to import EPS art
\caption{\label{fig:epsart}   Frequencies of individual waves which constitute the audio wave}
\end{figure}

\end{itemize}

In both of the instances above, initially we have the Amplitude vs. Time graph of the audio which shows just how many waves of different amplitudes have superposed to create the audio that we’re analyzing.  The amplitudes combine and change as a whole in different patterns with time and create a beautiful combination of individual waves.\\

Second, we have the frequency graph which is obtained once we apply the DFT on the original sound wave. The DFT picks out the peak frequency of individual waves within the original audio wave and plots them. Toward the left side of the graph, the lower frequencies pertain to the bass in the music while as we move toward the right, the higher frequencies pertain to the guitar and contributions of other musical instruments.\\

This is again very interesting because it precisely shows what the main frequencies which, in the form of sinusoids, act as building blocks to the audio are. It also shows how many frequencies are really important to us if we wanted to study the properties of the original audio. This operation of DFT helps us to modify the original audio as well because if know what frequencies are constitute the audio, we can also remove unwanted frequencies or modify and add others (this is explained further in the next section).

\subsection{Equalization using Eigen-matrices}

Equalization is a technique in music where amplitude of specific range of frequencies can be modified individually. For instance, in hip-hop songs, the low frequency bass is often amplified to give a bass boost that marks the signature music. Similarly, in Jazz and Orchestras, the middle as well as the high frequencies (also called treble) are signified more.\\

Doing such changes in a time dependent mixed audio signal (especially when all the instruments are recorded together) will be a tedious task by the process of simple addition. One would have to add the waves of the same exact shapes at those exact times, which would require a special solution in each case. However, since Fourier transform is a matrix which converts the basis from time to frequency domain, we can matrix-multiply it to more matrix and easily define a transform that can switch frequencies. Mathematically,

\begin{equation}
X=F A F^{-1}
\end{equation}

Here $A$ is a diagonal amplification matrix. Why a diagonal matrix? If we break the above three step transform, firstly we do the Fourier transform and get to the basis of frequencies which we simply want to scale (not shear or rotate). Once the appropriate frequencies are scaled (we can add new frequencies by scaling up and delete by scaling down). Then all we have to do is do the inverse transform which will give us back the transformed signal in time domain with the modified contribution of each frequency.\\

This form is the same as the Eigen matrix representation of any diagonalizable matrix. All the complications of different shapes and quality of sound waves (Timbre) is reduced down to sinusoids which are very easy to deal with. Each sinusoid is scaled by the respective element in the diagonal, giving us the equalized output.

\begin{figure}[hbt!]
\includegraphics[width=50mm]{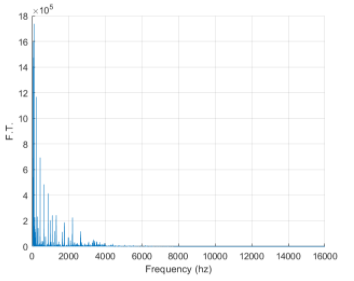}% Here is how to import EPS art
\caption{\label{fig:epsart}    Original Sample in frequency domain.}
\end{figure}

\begin{figure}[hbt!]
\includegraphics[width=50mm]{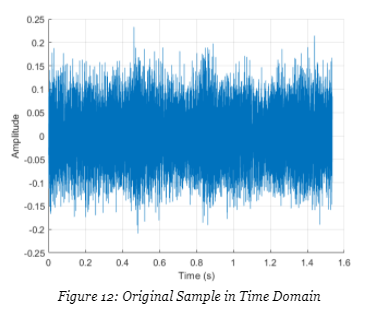}% Here is how to import EPS art
\caption{\label{fig:epsart}    Original Sample in time domain}
\end{figure}

\begin{figure}[hbt!]
\includegraphics[width=50mm]{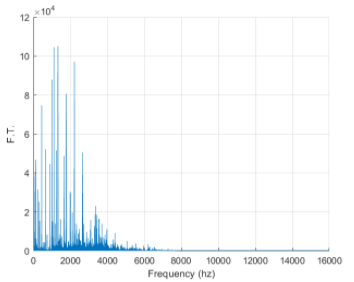}% Here is how to import EPS art
\caption{\label{fig:epsart}    High treble transformed sample (frequency)}
\end{figure}

\begin{figure}[hbt!]
\includegraphics[width=50mm]{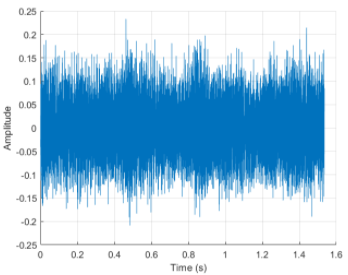}% Here is how to import EPS art
\caption{\label{fig:epsart}    High treble transformed sample (time)}
\end{figure}

\begin{figure}[hbt!]
\includegraphics[width=50mm]{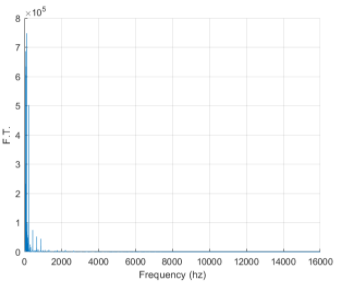}% Here is how to import EPS art
\caption{\label{fig:epsart}    Bass boost transformed sample  (frequency)}
\end{figure}

\begin{figure}[hbt!]
\includegraphics[width=50mm]{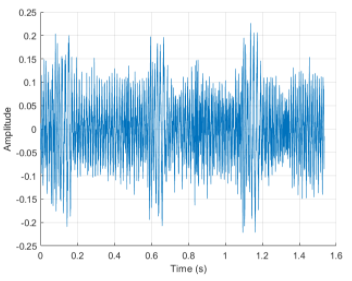}% Here is how to import EPS art
\caption{\label{fig:epsart}    Bass boost transformed sample  (time)}
\end{figure}

As one can see from figures 11 to 16, we have chosen a sample (1.5 sec sound wave from ‘Every Teardrop is a Waterfall’ by Coldplay) and then used the fast Fourier transform algorithm to perform the DFT. Then all we did was define a square diagonal matrix where each pivot represented the amplitude of the frequency corresponding in that row. For high treble (figures 13 and 14), the amplitudes were scaled by 0.1, 0.25, 0.5, 1, and 1, for ranges of frequencies from low to high respectively. As one can see, the initial frequencies (around 0-160hz, in our case) are scaled down by a huge amount which is clearly evident in the frequency domain, but less so in time domain. However, it can be observed that the time-domain wave is denser as the resultant frequency is higher (number of oscillations per sec will increase, making the wave closely packed together). Therefore, desired result is achieved.\\

In the case of bass boost, the opposite was done, which one can clearly see in the frequency domain where the relative peaks of low frequencies are higher than the high frequencies compared to that of the original wave. As a result, the time-dependent wave is less dense but has the same shape of wave hence not changing the audio itself.\\

We can see as the result that the shape of the wave is maintained throughout the transform and we do not observe any change of frequencies, yet we are able to amplify and de-amplify specific frequencies by such a transform. The Fourier transform did the tedious task of designing the wave while we just had to input the amplitude for individual frequencies. Due to this the timbre and the range of active frequencies (notes of musical instruments) is preserved, yet their amplitudes are changed independently.\\

We can see that this simple algorithm can be exploited further if we introduce shear and rotational transforms to it, therefore, giving us a lot of potential in non-trivial signal processing.

\section{Concluding Remarks}

This paper began with the understanding of discrete and continuous data and how continuous data can be converted to discrete data. It further explains the Fourier transform theoretically and the working of discrete Fourier transform in the field of Signal Processing with the help of computing. A code is built and implemented in MATLAB programming software which can process audio samples and yield interesting results, which opens a domain for huge variety of functions which can be applied on an audio sample.\\

Hopefully, the significance of the Fourier transform in signal processing, and more importantly, the beauty of linear algebra which aids the transform so well, was conveyed to the reader. We can take the advantage of converting continuous data into discrete, tangible vectors and apply these transforms as a consequence of linear algebra.
\nocite{*}

\bibliography{apssamp}% Produces the bibliography via BibTeX.

\end{document}